\documentclass{aa}
\usepackage{txfonts}
\usepackage{natbib}
\bibpunct{(}{)}{;}{a}{}{,}    
\usepackage{graphicx}
\usepackage{subfigure}
\usepackage[breaklinks=true, colorlinks=true, urlcolor=magenta, citecolor=blue, linkcolor=blue]{hyperref}

\begin{document}

\title{X-ray confirmation of the intermediate polar IGR~J16547-1916\thanks{Based on observations obtained with {\it NuSTAR} and {\it Swift}.}}

\author{
        A. Joshi\inst{\ref{WHU_1}} \and
        W. Wang\inst{\ref{WHU_1}, \ref{WHU_2}}\thanks{wangwei2017@whu.edu.cn} \and
        J. C. Pandey\inst{\ref{ARIES}} \and 
        K. P. Singh\inst{\ref{IISER}} \and
        S. Naik\inst{\ref{PRL}} \and
        A. Raj\inst{\ref{DU}}\and
        G. C. Anupama\inst{\ref{IIA}}\and
        N. Rawat\inst{\ref{ARIES}}
       }
       
\institute{
School of Physics and Technology, Wuhan University, Wuhan 430072, China \label{WHU_1}
\and
WHU-NAOC Joint Center for Astronomy, Wuhan University, Wuhan 430072, China\label{WHU_2}
\and
Aryabhatta Research Institute of observational sciencES, Manora Peak, Nainital 263 001, India\label{ARIES} 
\and
Department of Physical Sciences, Indian Institute of Science Education and Research Mohali, Sector 81, SAS Nagar, Manauli PO, 140306, India \label{IISER}
\and
Astronomy and Astrophysics Division, Physical Research Laboratory, Navrangpura, Ahmedabad, 380009, India\label{PRL}
\and
Department of Physics and Astrophysics, University of Delhi, 110007 Delhi, India\label{DU}
\and
Indian Institute of Astrophysics, Koramangala, Bangalore 560 034, India\label{IIA}
}

\abstract{%
Using X-ray observations from the {\it NuSTAR} and {\it Swift} satellites, we present temporal and spectral properties of an intermediate polar (IP) IGR J16547-1916. A persistent X-ray period at $\sim$ 546 s confirming the optical spin period obtained from previous observations is detected. The detection of a strong X-ray spin pulse reinforces the classification of this system as an intermediate polar. The lack of orbital or side-band periodicities in the X-rays implies that the system is accreting predominantly via a disk. A variable covering absorber appears to be responsible for the spin pulsations in the low energy range. In the high energy band, the pulsations are likely due to the self occultation of tall shocks above the white dwarf surface. The observed double-humped X-ray spin pulse profile indicates two-pole accretion geometry with tall accretion regions in short rotating IP IGR J16547-1916. We present the variation of the spin pulse profile over an orbital phase to account for the effects of orbital motion on the spin pulsation. X-ray spectra obtained from the contemporaneous observations of {\it Swift} and {\it NuSTAR} in the 0.5-78.0 keV energy band are modeled with a maximum temperature of 31 keV and a blackbody temperature of 64 eV, along with a common column density of 1.8$\times10^{23}$ cm$^{-2}$ and a power-law index of -0.22 for the covering fraction. An additional Gaussian component and a reflection component are needed to account for a fluorescent emission line at 6.4 keV and the occurrence of X-ray reflection in the system. We also present the spin phase-resolved spectral variations of IGR~J16547-1916 in the 0.5-78.0 keV energy band and find dependencies in the X-ray spectral parameters during the rotation of the white dwarf.
}


\titlerunning{X-ray confirmation of the intermediate polar IGR~J16547-1916}
\authorrunning{Joshi et~al.}
\maketitle

\begin{table*}
\caption{Log of the {\it Swift} and {\it NuSTAR} observations of IGR1654.\label{tab:obslog}}
\label{tab:Gaia_counterparts}
\centering
\begin{tabular}{@{}cccccccc@{}}
\hline\hline
Telescope    &Date of       &Observation  &Instrument      & Energy             &Integration      & Net Count Rate     \\
             & observation  &ID    &                & Band               &Time (s)         & (cts/s)      \\
\hline%
  {\it Swift}        & 21 Jan 2010  & 00090182001 & XRT            & 0.3-10.0 keV       &1568.02        &0.0938$\pm$0.0078  \\
             & 23 Jan 2010  & 00090182002 & XRT            & 0.3-10.0 keV       &3993.48    &0.0697$\pm$0.0043  \\
             & 13 Oct 2013  & 00040710001 & XRT            & 0.3-10.0 keV       &843.16             &0.0572$\pm$0.0083  \\
             & 25 Jan 2015  & 00040710002 & XRT            & 0.3-10.0 keV       &223.14             &0.1125$\pm$0.0225   \\
             & 29 Jun 2017  & 00040710003 & XRT            & 0.3-10.0 keV       &864.25             &0.0775$\pm$0.0096  \\
             & 30 Jun 2017  & 00040710004 & XRT            & 0.3-10.0 keV       &180.51             &0.1112$\pm$0.0249  \\
             & 08 Mar 2019  & 00088622001 & XRT            & 0.3-10.0 keV       &6461.66    &0.0862$\pm$0.0037  \\
             & 15 Mar 2019  & 00088622002 & XRT            & 0.3-10.0 keV       &4873.64    &0.0925$\pm$0.0045  \\
             & 17 Mar 2019  & 00088622003 & XRT            & 0.3-10.0 keV       &1476.58    &0.0629$\pm$0.0066  \\
  {\it NuSTAR}       & 16 Mar 2019  & 30460016002 & FPMA/FPMB      & 3-78 keV           &44565              &1.1100$\pm$0.0857  \\
\hline\hline
\end{tabular}
\end{table*}
\begin{figure*}
\centering
  \subfigure[]{\includegraphics[width=140mm]{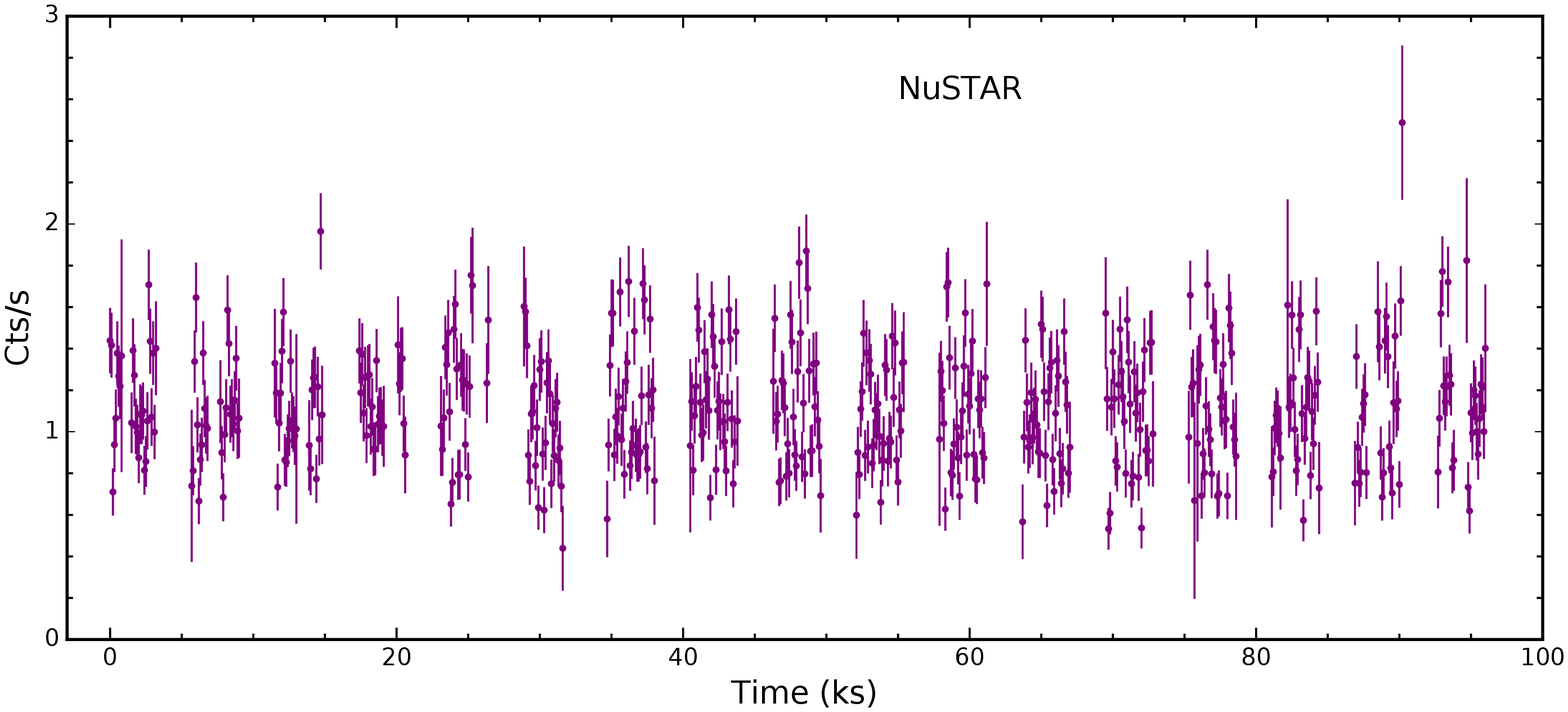}\label{fig:nuxraylc}}
  \subfigure[]{\includegraphics[width=140mm]{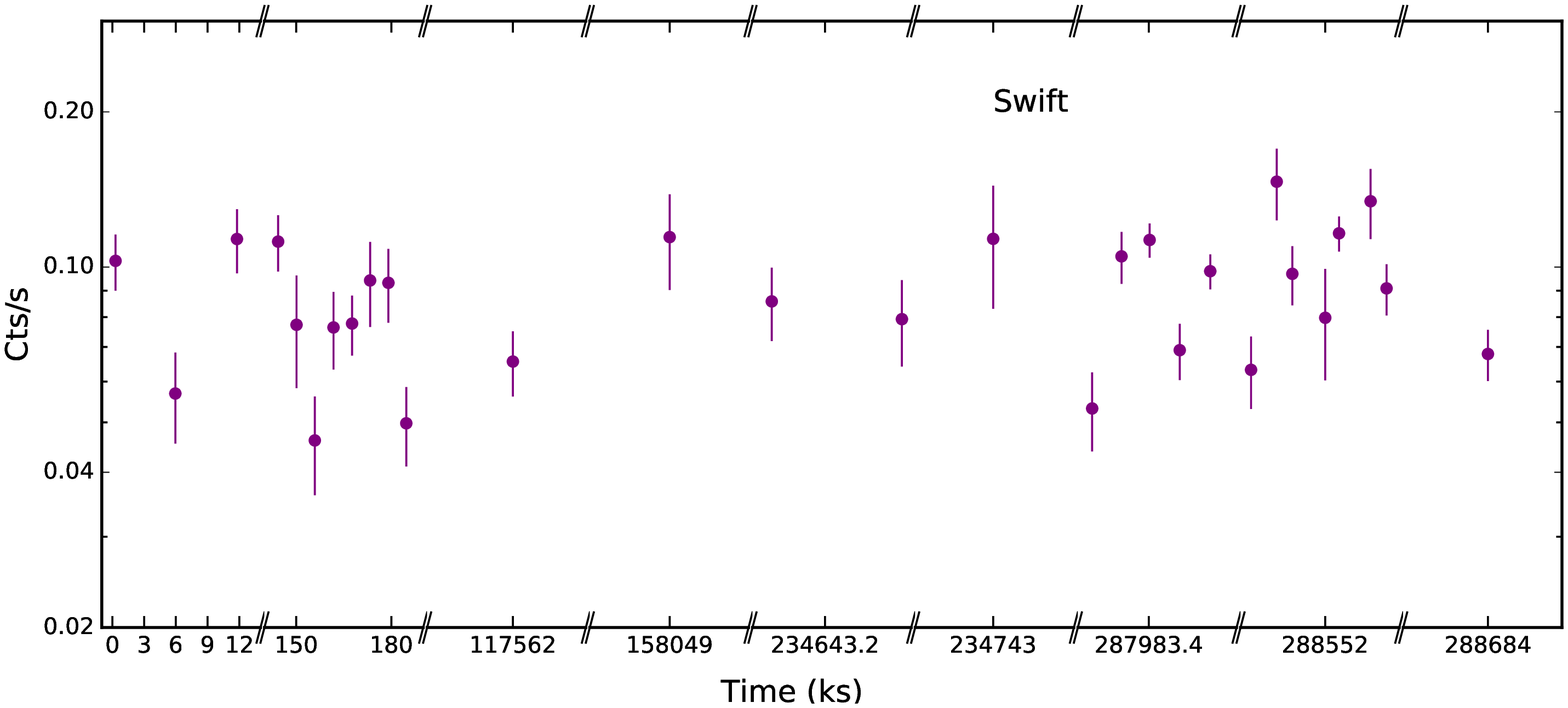}\label{fig:swxraylc}}
  \caption{X-ray light curves of IGR1654 obtained from (a) {\it NuSTAR} starting from MJD 58558.2323 and (b) {\it Swift} starting from MJD 55217.8178 with 100 s binning time in the 3-78 keV and 0.3-10.0 keV energy bands, respectively. The Swift-XRT light curves are grouped in each snapshot.}
\end{figure*}

\section{Introduction}
\label{sec:intro}
Magnetic cataclysmic variables (MCVs) possess a strong magnetic ($\sim$ 1-200 MG) white dwarf (WD) which accretes material from a Roche-lobe-filling red dwarf companion star. The intermediate polars (IPs), a subclass of MCVs, are known as asynchronized binaries (i.e., $P_\omega$ $<$ $P_\Omega$, where  $P_\omega$ and $P_\Omega$ are spin and orbital periods, respectively) with a WD magnetic field strength of $<$10 MG. The orbital-period distribution of MCVs shows that most of the IPs have orbital periods longer than the period gap of 2-3 hr \citep{Scaringi10}. In these systems, accreting material forms an accretion disk up to a certain point where the magnetic pressure exceeds the ram pressure and subsequently the accreting material flows along magnetic field lines. The accretion mechanism in IPs is usually explained with the three different scenarios viz disk-fed, disk-less, and disk-overflow \citep[see][]{Hameury86, Rosen88}. In these scenarios, the mode of accretion can be identified with the presence of orbital, spin, beat, and side-band frequencies. Along with these frequencies, various inner radii, such as the magneto-spheric radius ($R_{mag}$), the radius of the closest approach to the WD of a ballistic stream from the inner Lagrangian point ($R_{min}$), and the circularization radius ($R_{circ}$), can also be used to describe their accretion geometry \citep[for details, see][]{Warner95}.
From the sample of \cite{Pretorius14} and the Gaia DR2 parallaxes, a more updated space density of IPs is derived as $<$ 1.3$\times$10$^{-7}$ pc$^{-3}$ \citep[see][]{Schwope18}. With Gaia distances, it has been observed that the majority of the IPs peak at the hard luminosity of 10$^{33}$-10$^{34}$ erg s$^{-1}$, while there is also evidence of the existence of low-luminous IPs, mostly shorter-period systems below the period gap, and they have the luminosity of $\sim$ 10$^{30}$-10$^{32}$ erg s$^{-1}$ \citep{Schwope18, Martino20}. In the IPs, the magnetically channeled accretion column impacts the WD surface, and strong shocks are formed in the accretion columns; this heats the plasma up to a high temperature, about $\sim$ 50-600 MK. The shocked gas subsequently cools as it falls toward the surface of the WD via thermal bremsstrahlung emitting hard X-rays \citep{Aizu73}. However, some IPs also possess soft X-ray emission which can be modeled by using a blackbody component from the poles on the surface of the WD. Blackbody emission is mostly dominant in polars and has a temperature of 20-40 eV. In recent studies, a few IPs also possess a blackbody emission with a temperature more than that of the polars and such IPs are referred to as "soft-IPs" \citep[see,][]{Haberl95, Evans07}. The observed X-ray emission in IPs interacts with its surroundings and the part of X-ray emission undergoes photoelectric absorption and produces X-ray modulations. Around half of the intrinsic emission is also directed toward the WD and is expected to be reprocessed and/or reflected from the WD surface. The detection of the reflection can be confirmed with the two spectral features: a Compton reflection hump, which is observed in between the 10-30 keV energy range, and the presence of the fluorescent Fe K$\alpha$ emission line at 6.4 keV. Both absorption and reflection are important to carry out an in-depth study of an IP which requires broadband spectroscopy covering both soft and hard X-rays. 

\begin{figure*}
\centering
\subfigure[]{\includegraphics[width=70mm, height=78mm]{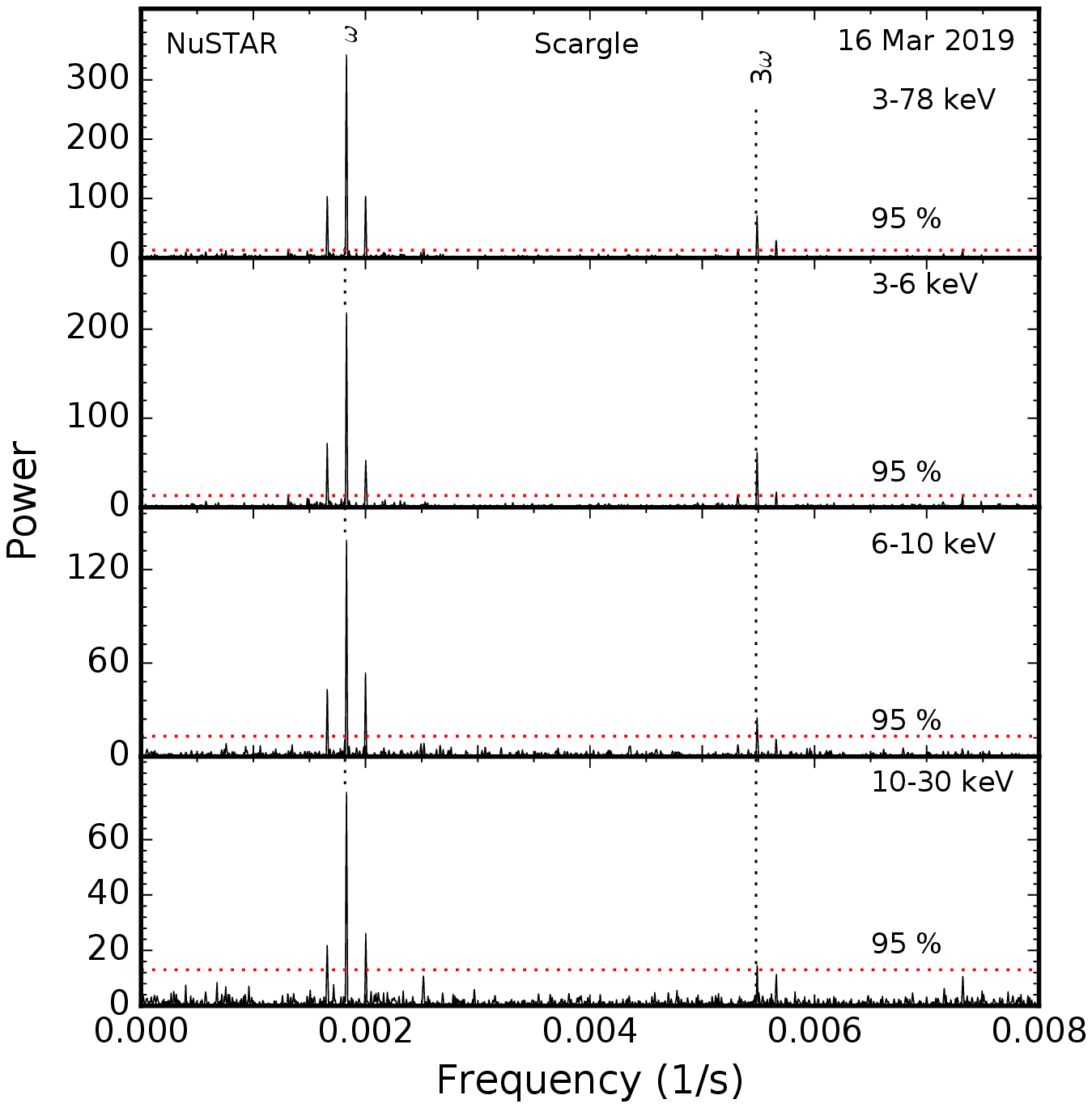}\label{fig:nuscps}}
\subfigure[]{\includegraphics[width=70mm, height=78mm]{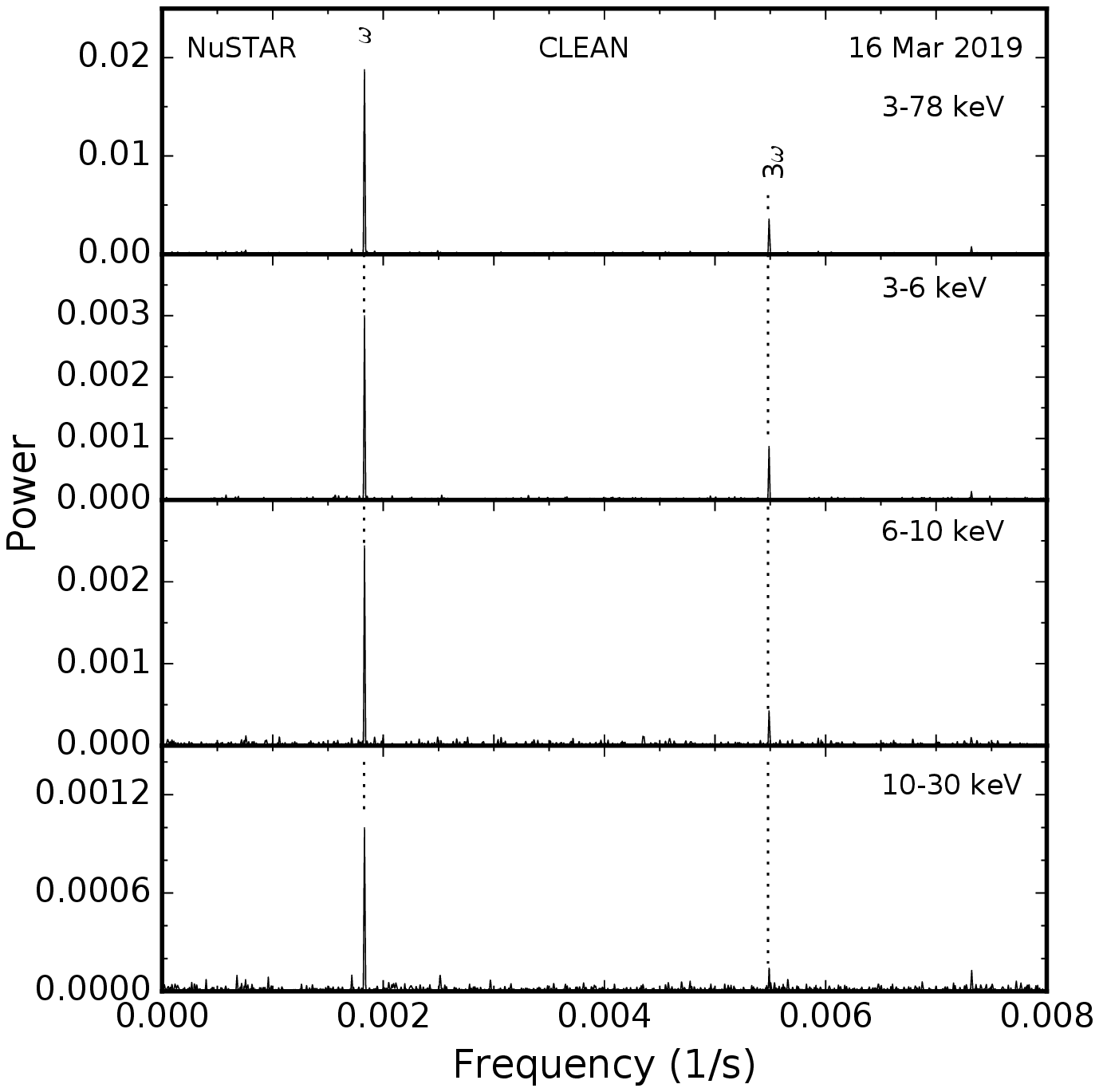}\label{fig:nuclps}}
\caption{(a) Lomb-Scargle and (b) CLEANed power spectra of IGR1654. From top to bottom, the panels show the X-ray power spectra obtained for the epoch 16 March 2019 of the {\it NuSTAR} observation in the 3-78 keV, 3-6 keV, 6-10 keV, and 10-30 keV energy bands, respectively. The red horizontal dotted lines represent the 95\% confidence level.}
\end{figure*}

\begin{table*}
\centering
  \caption{Periods corresponding to significant peaks in the power spectra of IGR1654 obtained from periodogram analysis of the {\it NuSTAR} and {\it Swift} data.}\label{tab:xrayps}
\setlength{\tabcolsep}{0.15in}
  \begin{tabular}{lcccccccccc}
\hline\\
    Telescopes                 &     Epoch & \multicolumn{2}{c}{Scargle } && \multicolumn{2}{c}{CLEAN} \\
\cline{3-4}\cline{6-7}\\
             &                 &    $P_{\omega}$        & P$_{3\omega}$  &&  $P_{\omega}$     &  P$_{3\omega}$    \\
             &                 &    ( s )               & ( s )          && ( s )             & (s)       \\
\hline\\
    {\it NuSTAR}       & 16 Mar 2019      &  546.3$\pm$0.8   & 182.2$\pm$0.1 && 546.3$\pm$0.8   & 182.2$\pm$0.1\\
    {\it Swift}        & 08 Mar 2019      &  546.4$\pm$3.0   & 182.1$\pm$0.3 && 545.1$\pm$0.3   & 182.1$\pm$0.1 \\
    {\it Swift}        & 15 and 17 Mar 2019   &  546.7$\pm$1.0   & ..            && 548.5$\pm$0.3   & ..            \\
    {\it Swift}        & 21 and 23 Jan 2010   &  546.6$\pm$0.4   & ..            && ..              & ..            \\
\hline                                                                                                         
\end{tabular}
\end{table*}

In this paper, we present the first detailed analysis of the X-ray observations of IGR~J16547-1916 (hereafter IGR1654). This source was detected in the {\it ROSAT} all-sky survey and cataloged in the {\it INTEGRAL}/IBIS survey \citep{Bird10}. \citet{Masetti10} tentatively classified this system as an IP based on the detection of strong Balmer and He II emission lines in its optical spectrum along with the equivalent width ratio of HeII  $\lambda (4686)$/H$\beta$ $\geq$ 0.5. From the optical variability, \cite{Lutovinov10} derived a period of 549$\pm$15 s and hypothesized this as a spin period of the WD. Later, \citet{Scaringi11} provided a clear detection of the orbital and spin periods of 3.7 hr and 546 s, respectively, using photometric and spectroscopic observations and classified this system as an IP. Recently, \cite{Shaw20} presented a legacy survey of 19 MCVs, including IGR1654 with the {\it NuSTAR}. They fit the {\it NuSTAR} spectra in the 20-78 keV energy band using the PSR X-ray spectral model \citep{Suleimanov16, Suleimanov19} and derived their WD masses. For IGR1654, they investigated the WD mass as 0.74$_{-0.08}^{+0.09}$ $M_\odot$. A detailed temporal and spectral analysis in the broad X-ray energy range has not been done for this source yet. We, therefore, present a detailed temporal and spectral analysis of IGR1654 using the {\it Swift} and {\it NuSTAR} observations. The paper is organized as follows: Section \ref{sec:obs} summarizes archival X-ray observations and their data reduction description. Analyses and the results of the X-ray data are described in Section \ref{sec:analysis}. Finally, we present a discussion and concise summary, in Sections \ref{sec:diss} and \ref{sec:sum}, respectively.

\begin{figure}
\centering
\includegraphics[width=92mm, height=105mm]{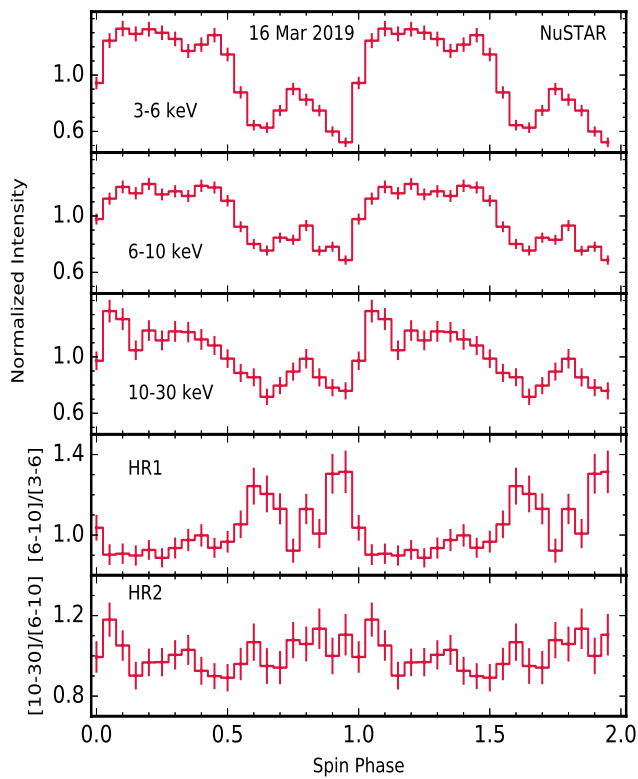}
  \caption{Folded X-ray light curves as obtained for the epoch 16 March 2019 of the {\it NuSTAR} observation of IGR1654 in the 3-6 keV, 6-10 keV, and 10-30 keV energy bands. The bottom panels show the hardness ratio curves HR1 and HR2, where HR1 is the ratio of the count rate in 6-10 keV to the count rate in the 3-6 keV energy bands, i.e., HR1=(6-10)/(3-6), and HR2 is the ratio of the count rate in 10-30 keV to the count rate in 6-10 keV energy bands, i.e., HR2=(10-30)/(6-10).} 
\label{fig:nuflc}
\end{figure}
\begin{figure}
\centering
\includegraphics[width=90mm]{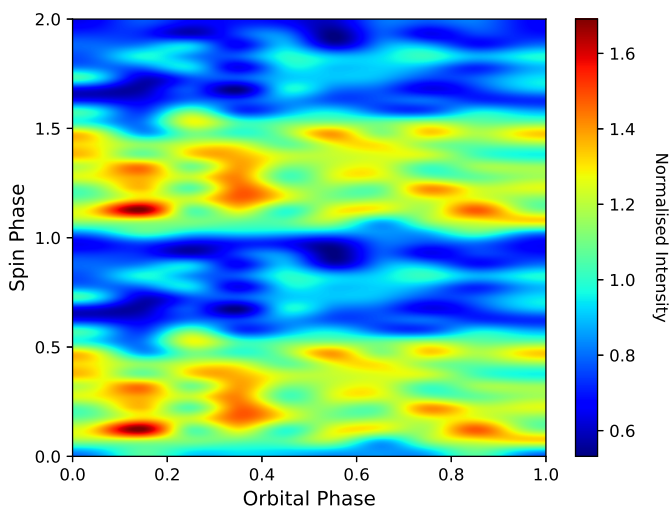}
  \caption{Orbital-phase-resolved spin pulse profile as obtained from the {\it NuSTAR} observations.}
\label{fig:orbspinprs}
\end{figure}
\begin{figure*}
\centering
  \subfigure[]{\includegraphics[width=90mm,height=95mm]{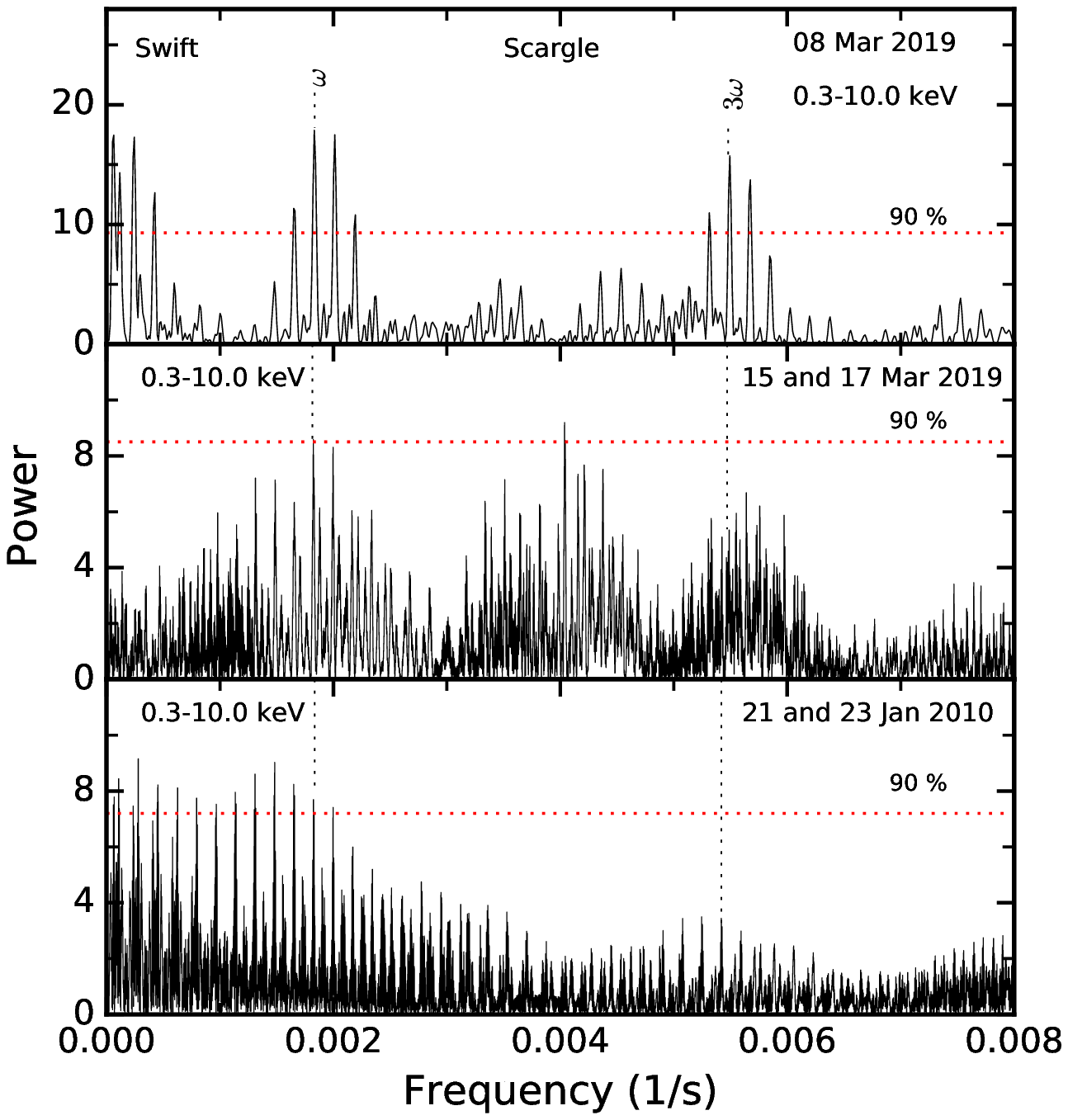}\label{fig:swscps}}
  \subfigure[]{\includegraphics[width=90mm,height=95mm]{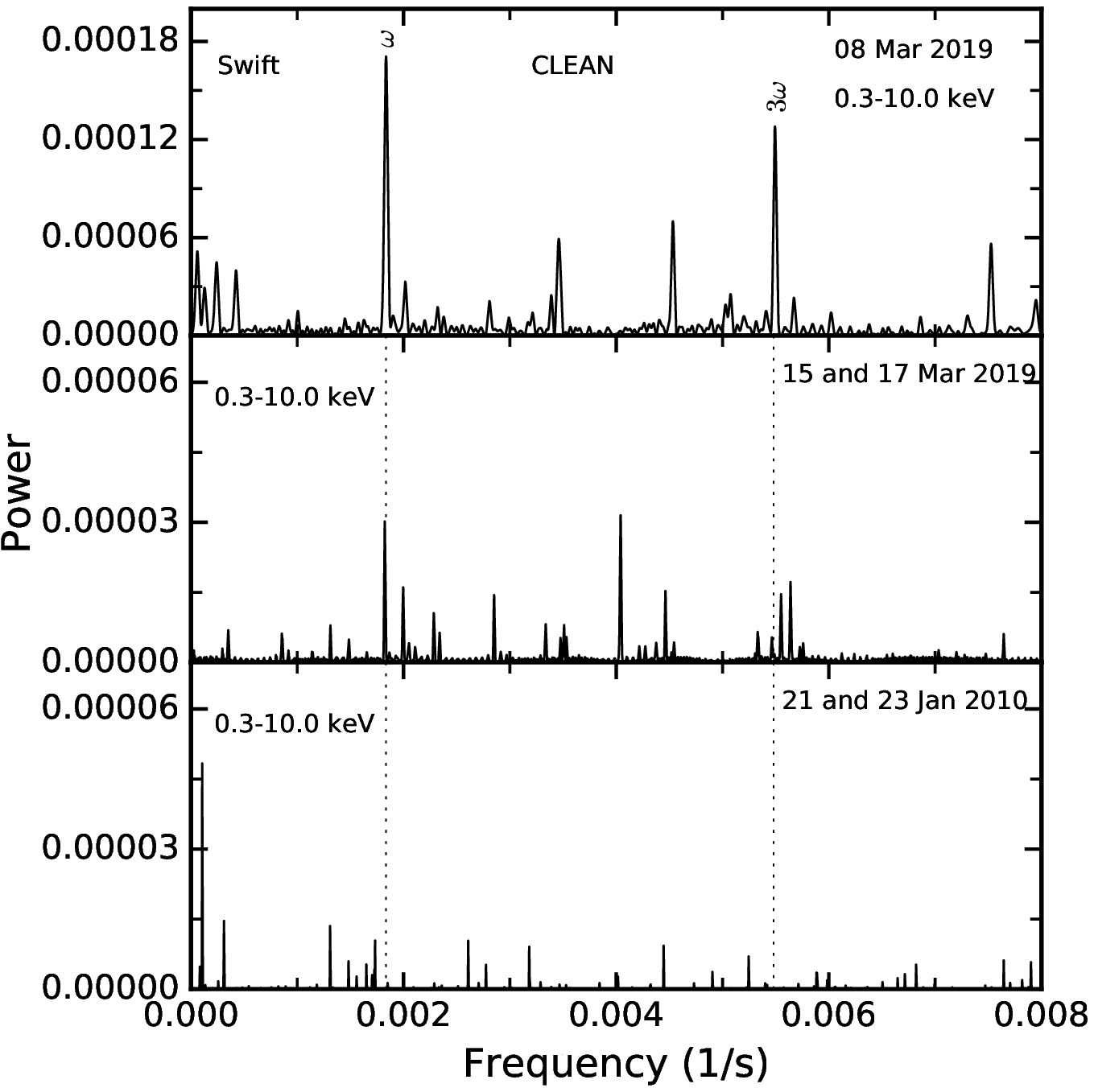}\label{fig:swclps}}
  \caption{(a) Lomb-Scargle and (b) CLEANed power spectra of IGR1654. From top to bottom, the panels show the X-ray power spectra obtained from the {\it Swift} observations for the epochs 08 Mar 2019, 15 and 17 Mar 2019, and 21 and 23 Jan 2010 in the 0.3-10.0 keV energy band, respectively. The red horizontal dotted lines represent the 90\% confidence level.} 
\end{figure*}
\begin{figure*}
\centering
\subfigure[]{\includegraphics[width=90mm,height=95mm]{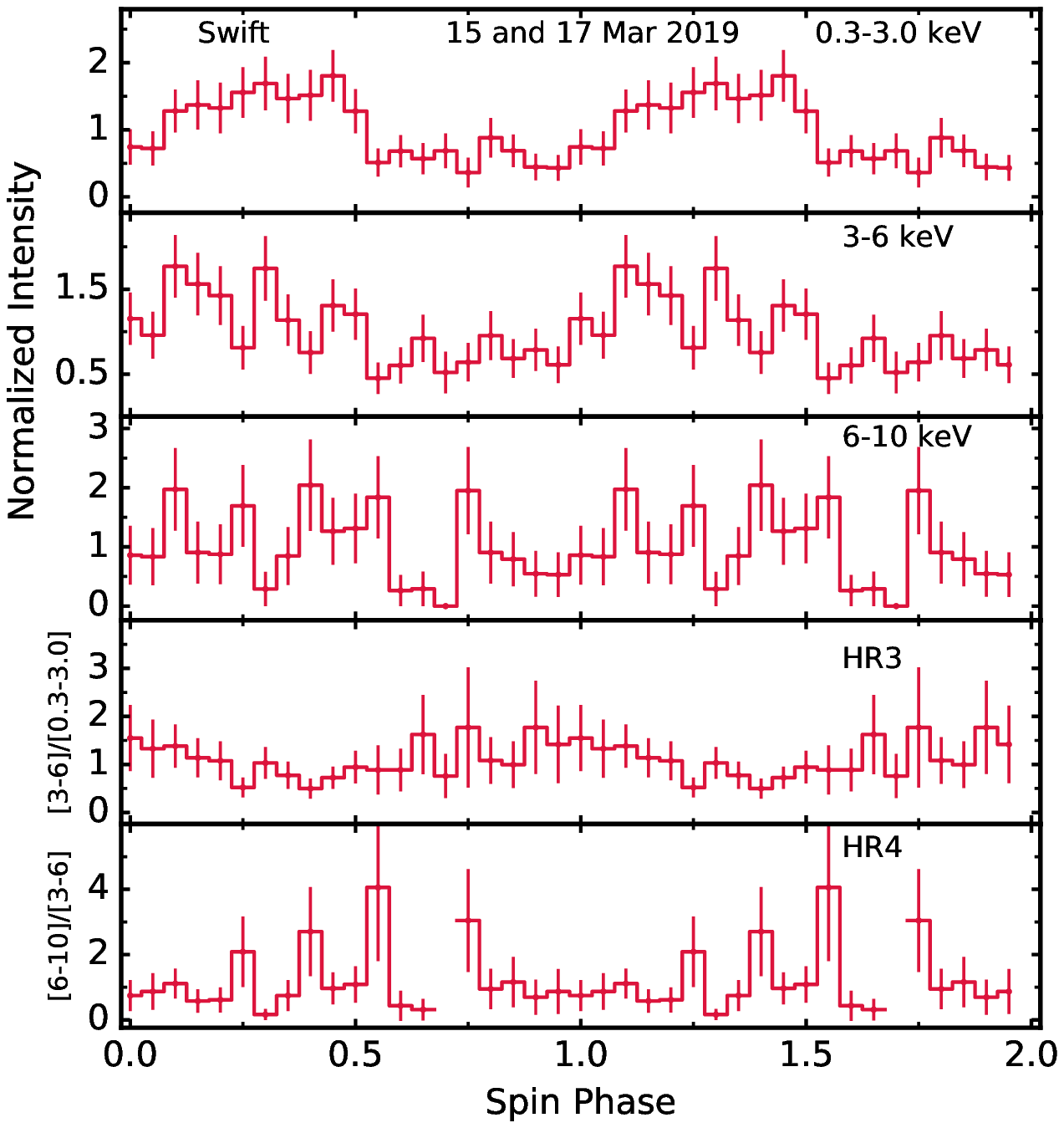}\label{fig:swflc15/17}}
\subfigure[]{\includegraphics[width=90mm,height=95mm]{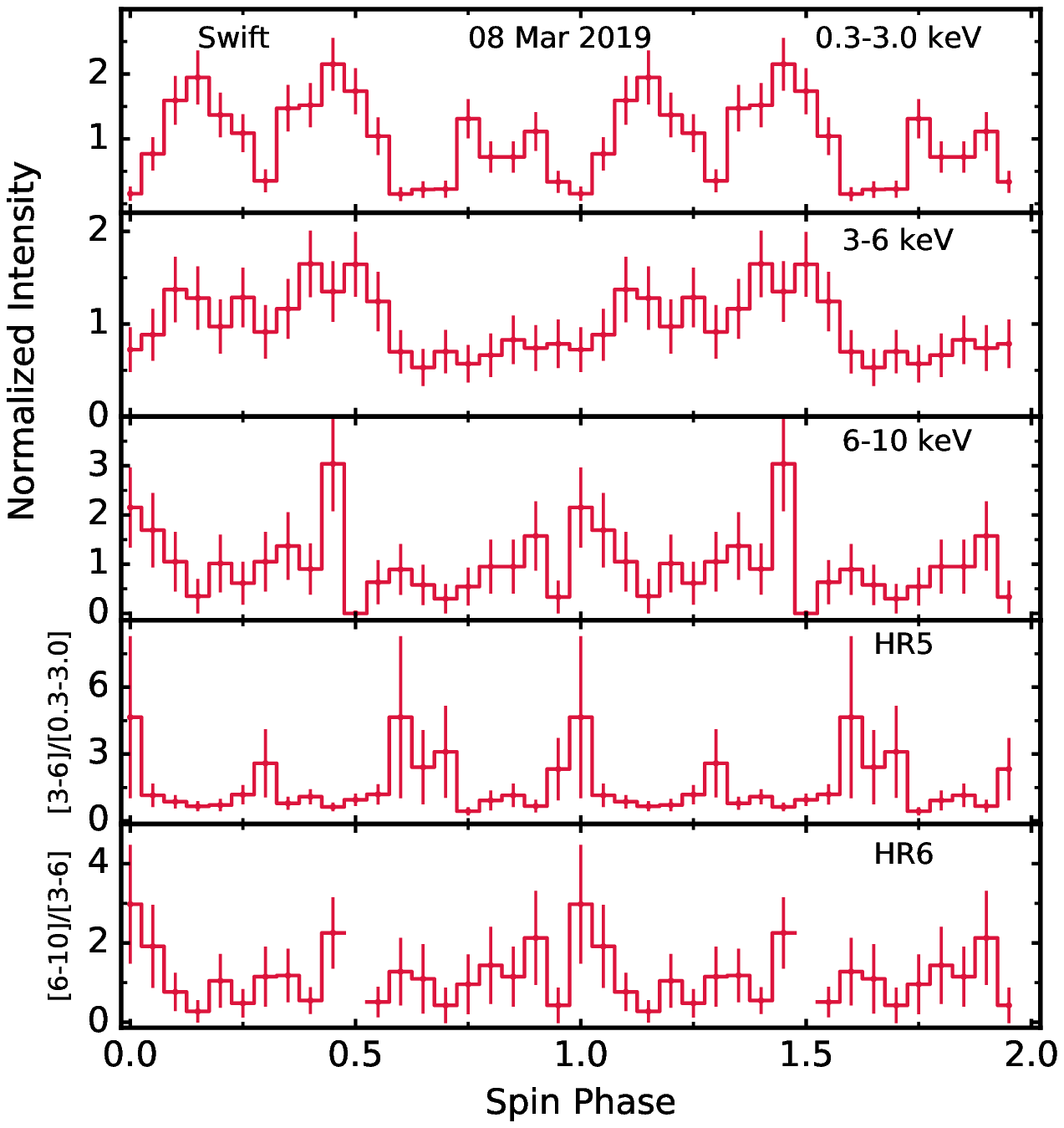}\label{fig:swflc08}}
  \caption{Folded X-ray light curves as obtained for the epochs 15 and 17 March 2019 and 08 March 2019 of the {\it Swift}-XRT observations of IGR1654 in the 0.3-3.0 keV, 3-6 keV, and 6-10 keV energy bands.  The bottom panels show the hardness ratio curves HR3, HR4, HR5, and HR6, where HR3 and HR5 are the ratio of the count rate in 3-6 keV to the count rate in 0.3-3.0 keV energy bands, i.e., HR3=HR5=(3-6)/(0.3-3.0), and HR4 and HR6 are the ratio of the count rate in 6-10 keV to the count rate in 3-6 keV energy bands, i.e., HR4=HR6=(6-10)/(3-6).}
\label{fig:swflc}
\end{figure*}


\section{Observations and data reduction}\label{sec:obs}
X-ray observations of IGR1654 were taken from the {\it NuSTAR} and {\it Swift} satellites and their log is given in Table \ref{tab:obslog}. We describe the details of the {\it NuSTAR} and {\it Swift} observations below.

\subsection{\it NuSTAR}
IGR1654 was observed with the hard X-ray focusing observatory {\it NuSTAR} \citep{Harrison13} on 16 March 2019 at 05:16:09 (UT) for 44.6 ks. {\it NuSTAR} consists of two co-aligned telescopes and two focal plane modules, FPMA and FPMB, and it observes in the 3-79 keV energy range. Each detector module consists of four 32$\times$32 Cadmium-Zinc-Telluride detectors and provides a spectral resolution (full width at half maximum, FWHM) of 0.4 keV at 10 keV and 0.9 keV at 68 keV. The data reduction was accomplished by using the standard {\it NuSTAR} Data Analysis Software (NuSTARDAS v1.4.1) as a part of  HEASOFT v6.26.1. The unfiltered events were first reprocessed by using {\sc nupipeline} in the presence of the updated version of calibration data files (CALDB 20191219) and then the science quality events were obtained after reprocessing. Source light curves and spectra were extracted by selecting a circular region of 70 arcsec around the source position. We used different extraction regions for FPMA and FPMB modules of the {\it NuSTAR}, based on respective images of each module, to consider the relative astrometric offset between them. For background light curves and spectra, the same size circular region located on the same detector chip and centered $\sim$ 4 arcmin away from the source was chosen to avoid contamination from the source photons. The barycentric corrected light curves, spectra, effective area files, and response matrices were obtained via the {\sc nuproducts} package. All spectra were grouped using {\sc grppha} to have at least 25 counts per bin. 

\subsection{{\it Swift}}
IGR1654 has been observed with the {\it Swift} satellite on nine occasions from 2010 to 2019. Among which, two observations on 15 March 2019 (ObsID 00088622002) and 17 March 2019 (ObsID 00088622003) are contemporaneous with {\it NuSTAR} observations. {\it Swift} consists of three instruments: the wide-field Burst Alert Telescope \citep[BAT;][]{Barthelmy05}, which covers the 15-350 keV energy range, and the narrow-field instruments, including the X-ray Telescope \citep[XRT;][]{Burrows05} which observes in the 0.3-10.0 keV energy range, and the UV/Optical Telescope \citep[UVOT;][]{Roming05} with filters covering 1700-6500 \AA. The task {\sc xrtpipeline} (version 0.13.4) along with the latest calibration files were used to produce the cleaned and calibrated event files. The barycentric correction was applied to the event files of all the observations using the task {\sc barycorr}. The source and background events with grades of 0-12 were extracted. The source product image, light curve, and spectrum were extracted by selecting a circular region of 30 arcsec radius. The background was chosen from  several nearby source-free regions with a similar size to that of the source. An ancillary response file (ARF) was also calculated  in order to correct the loss of the counts due to hot columns and bad pixels using exposure maps with the task {\sc xrtmkarf} and used the response matrix file (RMF), $swxpc0to12s6$\_2$0130101v014.rmf$, provided by the {\it Swift} team. All spectra from the {\it Swift/XRT} were rebinned using the {\sc grppha} for a minimum of five counts per bin. Spectral fits were performed by using the C-statistic for the fit, which is more suitable for low count per energy bin.


\section{Analysis and results} \label{sec:analysis}
\subsection{{\it NuSTAR}} \label{sec:nuanalysis}
\subsubsection{X-ray light curves and power spectra} \label{sec:nulcps}
Background-subtracted X-ray light curves of IGR1654 were obtained from the FPMA and FPMB instruments of {\it NuSTAR} and were combined using the FTOOLS task {\sc lcmath} \citep{Blackburn95}. The combined X-ray light curve in the 3-78 keV energy band is shown in Figure \ref{fig:nuxraylc}. To determine its periodic behavior, we performed a Fourier transform using the Lomb-Scargle periodogram (LS) method \citep{Lomb76, Scargle82, Horne86} from the light curves with a time bin of 10 s. Figure \ref{fig:nuscps} shows the Lomb-Scargle power spectra of the {\it NuSTAR} timing data in the 3-78 keV, 3-6 keV, 6-10 keV, and 10-30 keV energy bands. The LS power spectra in each energy band are dominated by two sets of signals $-$ one set centered around the frequency of 0.00183049 $s^{-1}$ and the other set around the frequency of 0.0054885 $s^{-1}$, which correspond to the frequencies $\omega$ and $3$$\omega$, respectively. In this way, we observed a total of six peaks in both sets of signals including $\omega$ and $3$$\omega$ (see Figure \ref{fig:nuscps}). We calculated the false alarm probability \citep{Horne86} to check the significance of detected peaks and found that all six peaks in the 3-78 keV energy band are more than 95\% of the confidence limit. The peak power of all these significant peaks is found to decrease toward the harder energy bands. Among all significant peaks, the strong power of $\sim$ 342 is found at frequency $\omega$, while a somewhat weaker peak with the power of $\sim$ 71 is detected at frequency $3$$\omega$ in the 3-78 keV energy band. Other side-band frequencies appear as the combination of spin and orbital frequencies, that is $\omega$$-$$2$$\Omega$, $\omega$$+$$2$$\Omega$ and $3$$\omega$$-$$2$$\Omega$, $3$$\omega$$+$$2$$\Omega$ on each side of $\omega$ and $3$$\omega$, respectively. These side-band frequencies are found to be equally spaced from the central frequencies $\omega$ and $3$$\omega$ with equal and unequal amplitudes, respectively.  However, their spacing does not correspond to the half of the orbital period of this system, which suggests that these side-bands might be spacecraft orbital aliases. Therefore, to further confirm the presence of true periodicities in the system, the light curve variations were modulated by the CLEAN algorithm \citep{Roberts87}. The CLEANed power spectra of IGR1654 in the 3-78 keV, 3-6 keV, 6-10 keV, and 10-30 keV energy bands are shown in Figure \ref{fig:nuclps}. The CLEANed power spectra were obtained with a loop gain of the 0.1 and 1000 iteration. In the CLEANed power spectra, we have not detected any side-band frequencies or unwanted spacecraft orbital aliases. However, only two strong peaks at frequencies $\omega$ and $3$$\omega$ were detected in each energy band, which is well consistent with the periods derived from the LS algorithm. The significant periods derived from the LS and CLEAN algorithms in the 3-78 keV energy band for the epoch 16 March 2019 are given in Table \ref{tab:xrayps}.

\subsubsection{Periodic intensity variations} \label{sec:nuenflc}
We explored the periodic variability of IGR1654, using the {\it NuSTAR} data, in the following three energy bands: 3-6 keV, 6-10 keV, and 10-30 keV ranges. The X-ray light curves in the aforementioned energy bands were folded using the spin period ephemeris reported by \cite{Scaringi11}. The orbital and spin period ephemerides of Scaringi likely have a cumulative error of well over one complete cycle by the time of the X-ray observations of the epoch 2019. In this case, the zero phase is not securely known, while more accurate periods are known. Therefore, we have considered a ``phase'' as a ``relative phase'' during binary and rotational motion throughout the paper. All {\it NuSTAR} light curves were folded with a phase bin of 0.05 and are shown in Figure \ref{fig:nuflc}. The spin-phase-folded light curves were found to be double-humped in all of the energy bands described above. The broad maxima or flat-topped hump was seen during the first half cycle; however, a relatively lower amplitude hump was observed during the subsequent half cycle. The spin modulations in each energy band are not sinusoidal, thus, we estimated the degree of spin pulsations with ($I_{max}$-$I_{min}$)/($I_{max}$+$I_{min}$) $\times$ 100 \%, where $I_{max}$ and $I_{min}$ are maximum and minimum intensities in a pulse profile, respectively. The derived value of the spin modulations are 43$\pm$3\% (3-6 keV), 28$\pm$3\% (6-10 keV), and 29$\pm$5\% (10-30 keV). The pulse profile was found to be more prominent at lower energies. During the rotation of the WD, we also extracted the hardness ratio curves (HR1 and HR2) between the hard and soft counts defined as follows: HR1 is the ratio of the count rate in [6-10] keV to the count rate in the [3-6] keV energy bands, that is, HR1=(6-10)/(3-6), and HR2 is the ratio of the count rate in [10-30] keV to the count rate in the [6-10] keV energy bands, that is, HR2=(10-30)/(6-10) and they are shown in the bottom two panels of Figure \ref{fig:nuflc}. The HR1 curve displays a strong modulation and is 180$^{\circ}$ out of phase with respect to the intensity modulation, that is, the maximum in the HR1 curve is observed at the lowest intensity. However, HR2 does not show any significant variation over the spin cycle. X-ray light curves of IGR1654 were also folded using the orbital period of 3.7 hr \citep{Scaringi11} in the 3-6 keV, 6-10 keV, 10-30 keV, and 3-78 keV energy bands. No orbital modulations in X-rays were noticed in these energy bands which was further confirmed by the absence of the orbital frequency in the {\it NuSTAR} power spectrum of IGR1654.

Using the {\it NuSTAR} observations, we also inspected the variation in the X-ray spin-pulse profile over an orbit of IGR1654. Using the orbital and spin period ephemerides reported by \cite{Scaringi11}, we extracted light curves for the 0.1 orbital phase interval in the 3-78 keV energy band and folded them with the spin period. We constructed an orbital-phase-resolved spin pulse profile that explores the evolution of pulse over an orbital cycle and is represented as the color composite plot (see Figure \ref{fig:orbspinprs}). A double or triple hump-like spin pulse profile appears to be present during an orbital motion of IGR1654. The shape and size of the pulse seem to vary during each orbital phase. Around the orbital phase segment 0.3-0.4, two humps appear to be merging in a single broad hump. Subsequently, the pulse shape again looks like a double or triple hump in forthcoming orbital phase intervals.

\begin{figure*}
      \centering
  \includegraphics[width=140mm]{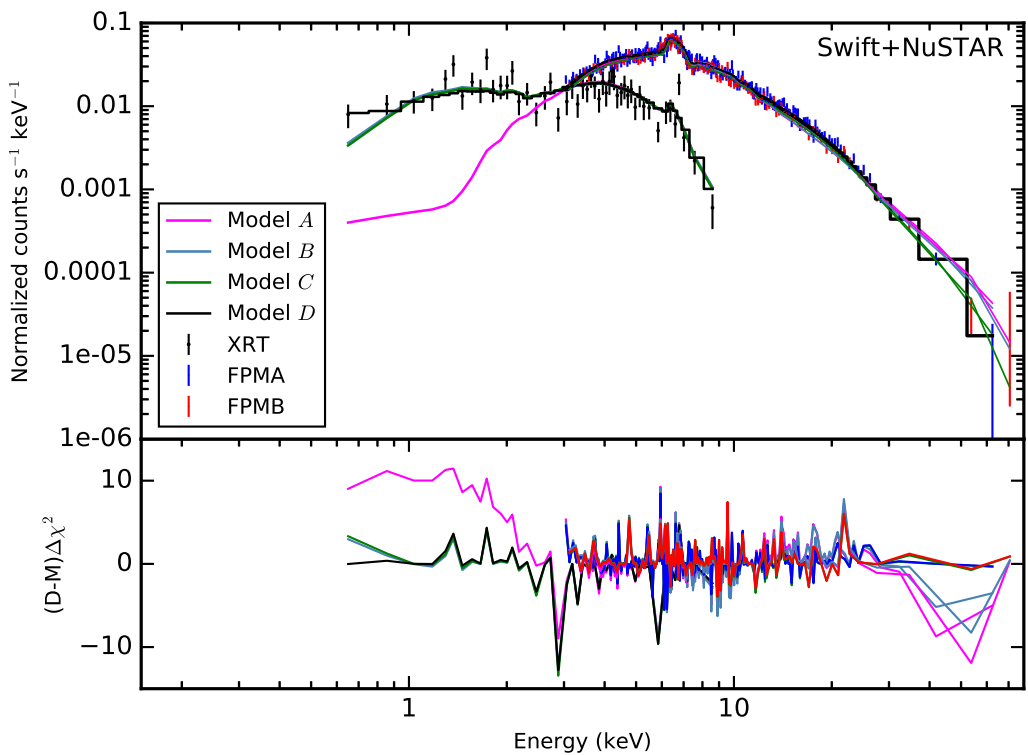}  
  \caption{Combined {\it Swift}-XRT (black), {\it NuSTAR}-FPMA (red), and {\it NuSTAR}-FPMB (blue) X-ray spectra of IGR1654 as fitted with the models $A$, $B$, $C$, and $D$ (see text for more details about these models). The bottom panel shows the spectral residuals obtained after fitting the broad-band spectrum with respect to all fitted models.}
\label{fig:xrayspec}
\end{figure*}

\subsection{{\it Swift}} \label{sec:swanalysis}
\subsubsection{X-ray light curves and power spectra} \label{sec:swlcps}
Figure \ref{fig:swxraylc} shows the X-ray light curve of IGR1654 in the 0.3-10.0 keV energy band as extracted for nine epochs of the {\it Swift}-XRT observations. Using the LS method, we performed a periodogram analysis of the {\it Swift} data. Due to large data gaps in the combined data set of the {\it Swift} observations, the real frequencies were found to be absent in the power spectrum. We, therefore, performed a power spectral analysis for each epoch separately from the light curves with a time bin of 10 s in the 0.3-10.0 keV energy band. A significant power spectrum was not obtained for the epochs 2013, 2015, and 2017 due to the small data coverage. However, we obtained the significant power spectra for the data of the epoch 08 March 2019 and a combined data set of the epochs 15, 17 March 2019 and 21, 23 January 2010. Figure \ref{fig:swscps} shows the LS power spectra of the timing data in the 0.3-10.0 keV energy band for the epochs described above. We detected two significant periods at frequencies $\omega$ and $3$$\omega$ for the epoch 08 March 2019, which are well consistent with the periods derived from the {\it NuSTAR} data. However, only a significant spin period was observed for the combined data set of the epochs 15, 17 March 2019 and 21, 23 January 2010. Although, the $3$$\omega$ frequency is also present in the power spectrum of these epochs, sadly they lie below the confidence level of 90\%. Similar to the {\it NuSTAR} data, the spacecraft orbital aliases are also present in the {\it Swift} LS power spectra of the epochs described above. Thus, to further confirm the presence of a true signal for these epochs, we employed the CLEAN algorithm to all of the epochs described above which are shown in Figure \ref{fig:swclps}. The CLEANed power spectra lead to the peaks at frequencies $\omega$ and $3$$\omega$ for the epoch 08 March 2019, while, only spin frequency $\omega$ was detected for the combined data set of the epochs 15 and 17 March 2019. Both detected frequencies are well consistent with the periods derived from their LS algorithm. However, no peaks at frequencies $\omega$ and $3$$\omega$ were detected in the CLEANed power spectrum of the combined data of the epochs 21 and 23 January 2010. The significant periods derived from the LS and CLEAN algorithms in the 0.3-10.0 keV energy band for the above mentioned epochs are given in Table \ref{tab:xrayps}.

\subsubsection{Periodic intensity variations} \label{sec:swenflc}
We folded the {\it Swift} X-ray light curves in the 0.3-3.0 keV, 3-6 keV, and 6-10 keV energy bands for the combined data of the epochs 15, 17 March 2019 and 08 March 2019. Figure \ref{fig:swflc} shows the folded X-ray light curves with a phase bin of 0.05 for the {\it Swift} data. Similar to the {\it NuSTAR}, for the epochs 15 and 17 March 2019 which are contemporaneous with {\it NuSTAR}, the {\it Swift} X-ray folded light curves seem to be double-humped in the 0.3-3.0 keV and 3-6 keV energy bands, where the second peak is marginally visible (see Figure \ref{fig:swflc15/17}). Using a similar approach as described in Section \ref{sec:nuenflc}, we also estimated the degree of spin pulsations as 67$\pm$25\%, 59$\pm$22\%, and 77$\pm$45\% in the 0.3-3.0 keV, 3-6 keV, and 6-10 keV energy bands, respectively. The rotational hardness ratio curves HR3 and HR4 are defined as the ratio of the count rates in [3-6] keV to the count rate in [0.3-3.0] keV, that is, HR3=(3-6)/(0.3-3.0), and the count rate in [6-10] keV to the count rate in [3-6] keV energy bands, that is, HR4=(6-10)/(3-6), respectively, were also derived and are shown in the bottom panels of Figure \ref{fig:swflc15/17}. Similar to the HR1 curve of the {\it NuSTAR} data, the HR3 curve shows hardening at the minimum intensity phase, and vice versa. However, no significant variation is seen in the HR4 curve. In contrast to the epochs 15, 17 March 2019 and 16 March 2019, the spin-phased {\it Swift} X-ray light curves in the 0.3-3.0 keV energy band reveals a triple-hump pulse profile for the epoch 08 March 2019 due to a stronger dip observed near phase 0.3 (see Figure \ref{fig:swflc08}). The first two humps seem to be prominent near the spin phases $\sim$ 0.15 and $\sim$ 0.45, while the third hump is observed with the smaller amplitude around spin phase $\sim$ 0.8. In 3-6 keV, a dip around 0.3 become significantly weaker and the modulation appears to be similar to {\it NuSTAR,} but the third peak after phase $\sim$ 0.6-0.7 seems to be negligible or marginally visible. The pulse amplitudes were derived as 87$\pm$24 \%, 51$\pm$21\%, and 75$\pm$44\% in the 0.3-3.0 keV, 3-6 keV, and 6-10 keV energy bands, respectively. The hardness-ratio curves, HR5 and HR6 (defined as similar to HR3 and HR4), were also extracted. The HR5 curve exhibits an anti-correlated pattern with the intensity profile, that is, at the maximum intensity phase, the hardness ratio is minimum (see bottom panel of Figure \ref{fig:swflc08}); however, the HR6 variation seems to be almost constant. The spin-phase-folded {\it Swift} X-ray light curves were also extracted for the epochs 2017, 2015, 2013, and 2010. No significant periodic variations were seen in the pulse profile of these epochs probably due to their insufficient data coverage. Similar to {\it NuSTAR}, the orbital-phase-folded {\it Swift} X-ray light curves were also explored for the epochs 21 and 23 January 2010, 08 March 2019, and 15 and 17 March 2019 in the 0.3-3.0 keV, 3-6 keV, 6-10 keV, and 0.3-10.0 keV energy bands. Unfortunately, their phase-coverage was found to be incomplete because the data length of {\it Swift} is small compared to the orbital period of the system.

\begin{table*}
\begin{center}
  \caption{Best-fit spectral parameters obtained from the spectral fitting of the contemporaneous {\it Swift}/XRT and {\it NuSTAR} (FPMA and FPMB) observations of IGR1654. \label{tab:xrayspec}}
\setlength{\tabcolsep}{0.05in}
  \begin{tabular}{ccccccccccccccccccc}
\hline\\
                                                                 & \multicolumn{2}{c}{pwab}                                  &   \multicolumn{2}{c}{reflect}         & \multicolumn{3}{c}{mkcflow}                             &&   \multicolumn{2}{c}{bbody}     &  \multicolumn{2}{c}{gauss}       &                         &                        \\
\cline{2-3} \cline{6-8} \cline{12-13}\\
               Models ($\downarrow$)                             &  N$_{H,max}$                         &  $\beta$                        &     $R_{refl}$ & Cos$i$         & $k$T  & $A_{z}$            & $n_{mkcflow}$                     &&  $kT_{bb}$   & $n_{bb}$                 &  $n_{g}$     &    EW     & $f_{X0.5-78.0}$                    & $\chi^{2}_{\nu}$/dof \\
\hline   \\                                                             
    A                                                            & ..                              & ..                       & ...                 & ...            &71$_{-7}^{+7}$  &1.0$_{-0.3}^{+0.3}$      &1.94$_{-0.19}^{+0.23}$      &&..                        & ..                           & 4.83$_{-0.59}^{+0.58}$    &  226$_{-21}^{+16}$          & 4.83$_{-0.05}^{+0.05}$      &1.24/792                \\~\\
    B                                                            & 2.1$_{-0.2}^{+0.5}$             & -0.32$_{-0.06}^{+0.05}$  & ...               & ...            &56$_{-3}^{+5}$  &1.0$_{-0.2}^{+0.2}$      &2.22$_{-0.18}^{+0.32}$      &&..                   & ..                           & 4.24$_{-0.55}^{+0.58}$    &  195$_{-12}^{+31}$          & 4.16$_{-0.04}^{+0.04}$      &1.13/791                \\~\\
    C                                                            & 1.8$_{-0.3}^{+0.3}$             & -0.26$_{-0.07}^{+0.08}$  & 1.7$_{-0.5}^{+0.6}$& 0.45$^\dagger$ &30$_{-3}^{+4}$  &0.6$_{-0.1}^{+0.1}$      &3.36$_{-0.28}^{+0.36}$      &&..                          & ..                           & 3.78$_{-0.57}^{+0.58}$    &  163$_{-31}^{+58}$          & 3.87$_{-0.04}^{+0.04}$       &1.05/790                 \\~\\
                                                                 & 2.0$_{-0.3}^{+0.3}$             & -0.26$_{-0.06}^{+0.06}$  & 1$^\dagger$        & $>0.70$        &30$_{-2}^{+3}$  &0.6$_{-0.1}^{+0.1}$      &3.51$_{-0.41}^{+0.20}$      &&..                          & ..                           & 3.89$_{-0.56}^{+0.56}$    &  181$_{-28}^{+34}$          & 3.97$_{-0.04}^{+0.04}$       &1.05/790                 \\~\\
    D                                                            & 1.8$_{-0.3}^{+0.3}$             & -0.22$_{-0.07}^{+0.10}$  & 1$^\dagger$        & $>0.70$        &31$_{-1}^{+3}$  &0.6$_{-0.1}^{+0.1}$      &3.43$_{-0.30}^{+0.31}$      &&64$_{-58}^{+47}$           & $<$ 1.22                     & 3.82$_{-0.55}^{+0.57}$    &  172$_{-15}^{+26}$          & 3.93$_{-0.04}^{+0.04}$      &1.04/788                 \\~\\
\hline
\end{tabular}
\end{center}
 {\it Notes:} ($\dagger$) represents a fixed parameter (see text for details). N$_{H,max}$ is the maximum equivalent hydrogen column in units of 10\textsuperscript{23} cm\textsuperscript{-2} and $\beta$ is the power-law index for the covering fraction. $R_{refl}$ is the reflection component, $Cosi$ is the inclination angle of the reflecting surface, $kT$ is the high temperature of {\sc mkcflow} model in units of keV, $n_{mkcflow}$ is the normalization of the {\sc mkcflow} model in units of $10^{-10}$ $M_\odot$ yr$^{-1}$, $kT_{bb}$ is the blackbody temperature in units of eV, $n_{bb}$ is the normalization constant of the blackbody in units of $10^{-2}$, $n_{g}$ is normalization constant of the Gaussian component in units of 10\textsuperscript{-5}, EW is the equivalent width of Fe K$\alpha$ in units of eV, and $f_X$ is the unabsorbed X-ray flux derived in the 0.5-78.0 keV energy band in units of 10\textsuperscript{-11}erg cm\textsuperscript{-2} s\textsuperscript{-1}. All the errors are within a 90\% confidence interval for a single parameter ($\Delta$ $\chi^2$ =2.706).\\
\end{table*}

\begin{table*}
\begin{center}
  \caption{Best-fit spectral parameters derived from the {\it Swift}/XRT observations. \label{tab:swiftspecparam}}
\setlength{\tabcolsep}{0.02in}
\begin{tabular}{ccccccccccc}
  \hline \\ 
                & \multicolumn{2}{c}{pwab}       & mkcflow    &   \multicolumn{2}{c}{bbody}      &      &                         \\
\cline{2-6} \\
            {\it Swift}/XRT     & N$_{H,max}$          &  $\beta$                  & $n_{mkcflow}$           & $kT_{bb}$     & $n_{bb}$   & $f_{X0.5-10.0}$              & $\chi^{2}_{\nu}$/dof \\
          (Date of Obs.) & (10$^{23}$ cm\textsuperscript{-2})          &            & ($\times$10\textsuperscript{-10} $M_\odot$ yr$^{-1}$) & (eV) & ($\times$$10^{-02}$)   & ($\times$10$^{-11}$ erg cm\textsuperscript{-2} s\textsuperscript{-1}) &                          \\
\hline \\
    21 and 23 Jan 2010  & 2.5$^{+4.8}_{-1.2}$        & -0.21$^{+0.17}_{-0.13}$ & 4.01$^{+2.59}_{-0.98}$   & $<$ 64              & $<$ 3.2               & 4.63$^{+0.40}_{-0.40}$         & 1.01/75  \\~\\
    08 Mar 2019     & 2.0$^{+4.8}_{-1.0}$        & -0.35$^{+0.17}_{-0.12}$ & 3.00$^{+1.78}_{-0.63}$   & 53$^{+6}_{-11}$     & $<$ 2.6               & 3.60$^{+0.30}_{-0.30}$         & 1.00/96   \\~\\ 
\hline
\end{tabular}
\end{center}
\end{table*}

\begin{table*}
\begin{center}
  \caption{Best-fit spectral parameters derived from the spectral fitting of the contemporaneous {\it Swift}/XRT and {\it NuSTAR} (FPMA and FPMB) observations of IGR1654 at spin phases. \label{tab:xrayspinprs}}
\setlength{\tabcolsep}{0.15in}
\begin{tabular}{ccccccccccccccccccc}
\hline\\
Model        & Parameters & \multicolumn{5}{c}{Spin Phase}\\~\\
\cline{3-7}\\ 
                     &    &  0.0$-$0.1      &   0.1$-$0.2   &  0.2$-$0.3     & 0.3$-$0.4 & 0.4$-$0.5 \\                                                                     
\hline\\
 pwab        & $N_{H,max}$ ($\times$10\textsuperscript{23} cm\textsuperscript{-2})                                & 1.26$_{-0.36}^{+0.59}$    & 0.99$_{-0.36}^{+0.59}$      & 1.73$_{-0.47}^{+0.75}$        & 2.79$_{-0.81}^{+1.33}$            &1.41$_{-0.43}^{+0.59}$           \\ 
                     & $\beta$                                                                                             & $>$ -0.55                 & -0.05$_{-0.35}^{+0.59}$     & -0.26$_{-0.15}^{+0.22}$       & -0.38$_{-0.15}^{+0.21}$           &-0.30$_{-0.17}^{+0.29}$          \\ 
Mkcflow    & $n_{mkcflow}$ ($\times$10\textsuperscript{-10}) $M_\odot$ yr$^{-1}$                                  & 3.57$_{-0.78}^{+0.75}$    & 3.73$_{-0.61}^{+0.36}$      & 4.20$_{-0.29}^{+0.33}$        & 4.38$_{-0.29}^{+0.34}$            &3.83$_{-0.26}^{+0.29}$           \\
Gaussian   & $n_g$ ($\times$10\textsuperscript{-5})                                                              & 2.25$_{-1.51}^{+1.56}$    &4.37$_{-1.76}^{+1.68}$       & 3.93$_{-1.78}^{+1.87}$        & 3.33$_{-1.87}^{+1.99}$            &5.00$_{-1.74}^{+1.83}$           \\
X-ray flux & $f_{X0.5-78.0}$($\times$10\textsuperscript{-11}erg cm\textsuperscript{-2} s\textsuperscript{-1})  & 3.98$_{-0.11}^{+0.11}$    &4.19$_{-0.12}^{+0.11}$       & 4.70$_{-0.13}^{+0.13}$        & 4.90$_{-0.14}^{+0.13}$            &4.32$_{-0.12}^{+0.12}$            \\
                     & $\chi_\nu^2$/(dof)                                                                                  & 1.26/(140)                 &1.10/(147)                    & 1.08/(151)                     & 0.96/(147)                         &1.08/(150)                       \\
                     &  & \\
\cline{3-7}\\ 
                     &    &  0.5$-$0.6      &   0.6$-$0.7   &  0.7$-$0.8     & 0.8$-$0.9 & 0.9$-$1.0 \\                                                                     
\hline\\
pwab        & $N_{H,max}$ ($\times$10\textsuperscript{23} cm\textsuperscript{-2})                                & 2.02$_{-0.64}^{+1.04}$    & $>$ 6.22                    & 1.99$_{-0.73}^{+1.16}$        & 2.84$_{-1.08}^{+2.27}$            &4.04$_{-1.58}^{+2.55}$           \\ 
                     & $\beta$                                                                                             & 0.01$_{-0.31}^{+0.79}$    & -0.62$_{-0.10}^{+0.15}$     & -0.11$_{-0.33}^{+0.53}$       & -0.25$_{-0.26}^{+0.42}$           &-0.26$_{-0.21}^{+0.51}$            \\ 
Mkcflow    & $n_{mkcflow}$ ($\times$10\textsuperscript{-10}) $M_\odot$ yr$^{-1}$                                  & 3.07$_{-0.41}^{+0.37}$    & 3.67$_{-0.43}^{+0.68}$      & 3.05$_{-0.45}^{+0.10}$        & 2.89$_{-0.33}^{+0.38}$            &2.93$_{-0.32}^{+0.36}$           \\
Gaussian   & $n_g$ ($\times$10\textsuperscript{-5})                                                              & 3.68$_{-1.53}^{+1.77}$    &5.65$_{-2.58}^{+3.18}$       & 2.60$_{-1.50}^{+1.47}$        & 5.52$_{-1.79}^{+1.95}$            &4.80$_{-1.98}^{+2.08}$           \\
X-ray flux & $f_{X0.5-78.0}$($\times$10\textsuperscript{-11}erg cm\textsuperscript{-2} s\textsuperscript{-1})  & 3.45$_{-0.11}^{+0.11}$    &4.14$_{-0.15}^{+0.14}$       & 3.42$_{-0.11}^{+0.11}$        & 3.27$_{-0.12}^{+0.11}$            &3.30$_{-0.12}^{+0.12}$            \\
                     & $\chi_\nu^2$/(dof)                                                                                  & 0.82/(114)                 &0.90/(96)                     & 0.99/(107)                     & 0.90/(99)                          &1.05/(90)                          \\~\\ 
\hline
\end{tabular}
\end{center}
\end{table*}

\begin{figure}
\centering
\includegraphics[width=90mm, height=92mm]{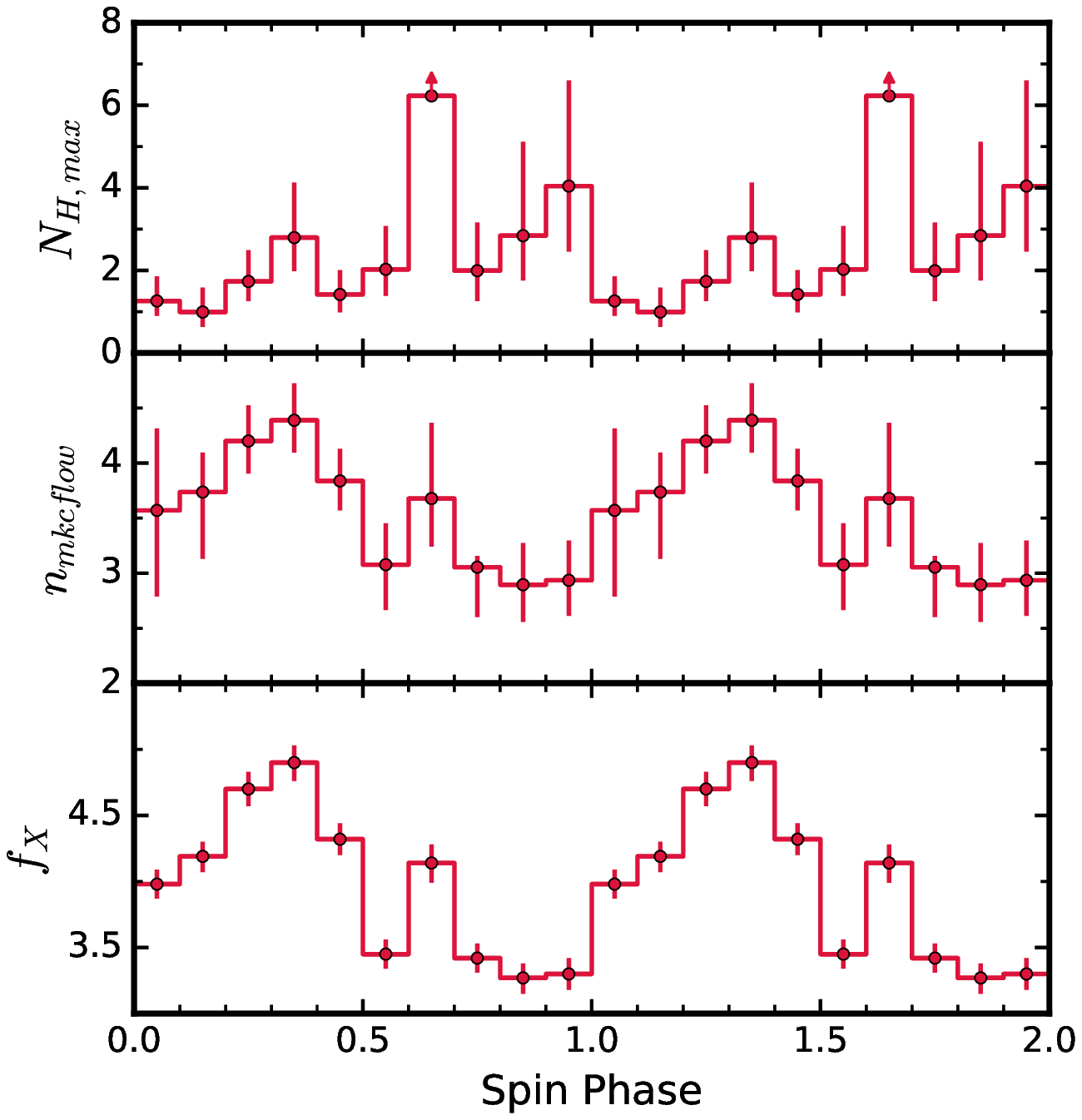}
\caption{Variations of the best-fit spectral parameters as a function of spin phases. From top to bottom, the panels show the variation in the maximum equivalent hydrogen column $N_{H,max}$, the normalization of the {\sc mkcflow} model ($n_{mkcflow}$), and the unabsorbed X-ray flux ($f_{X}$) in the 0.5-78.0 keV energy band. The unit of all parameters is similar as described in Table \ref{tab:xrayspec}. Error bars are plotted within a 90\% confidence limit for a single parameter.}
\label{fig:xrayspinprs}
\end{figure}


\subsection{X-ray spectra} \label{sec:xrayspec}
\subsubsection{{\it Swift} and {\it NuSTAR} spectral fits} \label{sec:simspec}

Background-subtracted X-ray spectra of IGR1654 are shown in Figure \ref{fig:xrayspec}. The iron line complex is present in both the {\it NuSTAR} and {\it Swift} spectra. The X-ray spectral analysis was performed using XSPEC version-12.10.1 \citep{Arnaud96, Dorman01}. The spectral fitting of the {\it Swift} observations was performed in the 0.5-10.0 keV energy range due to poor signal-to-noise ratio of individual {\it Swift} spectra below 0.5 keV. The spectral analysis of the {\it NuSTAR} data was done in the 3-78 keV energy band. We investigated IGR1654 by the spectral fitting of the contemporaneous {\it Swift}/XRT (15 March 2019 and 17 March 2019) and {\it NuSTAR}-FPMA/FPMB (16 March 2019) observations. To fit the broad 0.5-78.0 keV spectra, we used a multi-temperature cooling flow model {\sc {mkcflow}} \citep{Mushotzky88} which is more appropriate for the accreting WDs \citep[see][]{Mukai17} along with a Gaussian component with the fixed-line energy and line width at 6.4 keV and 0.01 keV, respectively, and approximated with the interstellar absorption model {\sc phabs} using the abundance tables of \cite{Asplund09} and the photoelectric absorption cross-sections {\sc BCMC} \citep{Balucinska92}. To know the cross-calibration uncertainties of the distinct instruments, a constant model component was also included as model A=constant$\times$phabs(mkcflow + gauss). The abundances were applied in the optically thin thermal model from \cite{Asplund09}. The redshift required in the mkcflow model cannot be zero. It was thus fixed to a value of 17.766$\times$$10^{-8}$ for a Gaia distance of 1066$_{-54}^{+61}$ pc \citep{GaiaCollaborationBailejonesr18} and a cosmological Hubble constant of 50 km s$^{-1}$ Mpc$^{-1}$. The low temperature of the {\sc {mkcflow}} was fixed to the minimum value allowed by the model as 0.0808 keV. We assumed a switch parameter at the value of 2 which determines whether the spectrum is computed by using the AtomDB data. With model $A$, we observed an equivalent hydrogen column (NH) of 8$\times$$10^{22}$ cm$^{-2}$ from the {\sc phabs} model, which is very large at the Gaia distance of this source. A fully covering absorber model {\sc phabs} is not necessarily due to interstellar absorption entirely $-$ a component of an absorber intrinsic to the IP may well have the same energy dependence as phabs. In that case, the intrinsic absorption might be affected due to the dominant local absorbers, resulting in a high value of NH. Moreover, the spectrum did not fit  at lower energies below 2 keV well (see Figure \ref{fig:xrayspec}) with $\chi^2_\nu$ of 1.24 (792 dof) using a phabs absorption model, which confirms the presence of the complex absorption. Thus, to account for the complex absorption, we used the {\sc pwab} model as B=constant$\times$phabs$\times$pwab(mkcflow + gauss) which is a power-law distribution of a covering fraction as a function of the maximum equivalent hydrogen column N$_{H,max}$ and the power-law index for the covering fraction $\beta$ \citep{Done98}. Using model B, we found a low value of NH $\sim$ 10$^{18}$ cm$^{-2}$ from {\sc phabs} which is not physical. Thus, for further spectral fitting, we fixed the NH value to the total Galactic column in the direction of IGR1654, that is, at 1.27$\times$$10^{21}$ $cm^{-2}$ \citep{Kalberla05}. Model B with a fixed NH provides a better fit to the spectra below 2 keV with an improved $\chi^2_\nu$ of 1.13 (791 dof). Using model B, we obtained the large equivalent width (EW) of the Fe K$\alpha$ emission line, and the column density of cold matter was found close to the value of 2$\times$$10^{23}$ cm$^{-2}$ \citep[see][]{Inoue85}. This suggests that the bulk of the Fe K$\alpha$ emission line can be accounted by the absorbers, while a small contribution, if any, is from reflection. Using models A and B, we also found that the spectrum did not fit at higher energies well, which could be associated with the absence of a reflection component. Therefore, to take the occurrence of X-ray reflection in the system into account, we used a convolution model {\sc {reflect}} \citep{Magdziarz95} as model C=constant$\times$phabs$\times$pwab(reflect$\times$mkcflow + gauss). With model C, we found a better fit, but the value of the reflection scaling factor was derived to be $\sim$ 1.7 (more than unity) at a fixed inclination angle of the reflecting surface at the default value of cos $i$=0.45. The reflection component describes the fraction of downward radiation that is reflected, so, the observed value more than unity is not physical. Thus, we fixed this parameter to unity, allowing cos $i$ to vary, and we obtained fits with a similar quality as the best fit value cos $i$ > 0.70. The inclusion of the {\sc {reflect}} model provides a more satisfactory fit at higher energies with an improved $\chi^2_\nu$ of 1.05 (790 dof). However, model C was unable to explain the positive residual below 1 keV, which is visible in the residual plot of Figure \ref{fig:xrayspec}. To account for this excess in soft X-rays, we fit the spectra with a lower-temperature optical thin plasma component \citep[{\sc APEC};][]{Smith01}, but this component adjusted the values of other parameters such that the spectral fitting gave an unacceptable fit. After that, to account for the soft energy residual pattern, an additional blackbody component was then added as model $D$=constant$\times$phabs$\times$pwab(reflect$\times$mkcflow+bb+gauss), which provides a better fit to the spectra with a marginally improved $\chi^2_\nu$ of 1.04 (788 dof). The F-test showed that the model D was more significant with an F-statistics of 4.8 with a null hypothesis probability of $8.2 \times 10^{-3}$.  However, the confinement of the soft X-ray excess to only one spectral bin reduces its evidence in the IGR1654. Unabsorbed X-ray flux in the 0.5-78.0 keV energy band was also calculated using the ``cflux" model. The best-fit parameters as obtained for each model are given in Table \ref{tab:xrayspec}, where the error bars are quoted with a 90\%\ confidence limit for a single variable parameter.

\subsubsection{{\it Swift}/XRT spectral fits}
We examined the spectral evolution of IGR1654 using other {\it Swift}/XRT observations. The same best-fit model D as described in the Section \ref{sec:simspec} was used for the spectral fitting of the {\it Swift}/XRT observations, keeping fixed the plasma temperature and abundance values of the {\sc mkcflow} and the parameter of the Gaussian component to those determined from the average spectral fitting of the {\it NuSTAR} and {\it Swift} observations as presented in Table \ref{tab:xrayspec}. We also used the same fixed parameters which were adopted for the average spectral fitting. For the epochs 2013, 2015, and 2017, the spectra were found to be under-fitted as per their low value of reduced chi-square. Thus, due to their poor statistics, we did not include the data of these epochs further for spectral analysis. However, we estimated their fluxes using {\sc webpimms} \footnote{\textcolor{blue} {{https://heasarc.gsfc.nasa.gov/cgi-bin/Tools/w3pimms/w3pimms.pl}}} as $\sim$ 3.4$\times$$10^{-12}$ erg cm$^{-2}$ s$^{-1}$, $\sim$ 6.7$\times$$10^{-12}$ erg cm$^{-2}$ s$^{-1}$, and $\sim$ 5.6$\times$$10^{-12}$ erg cm$^{-2}$ s$^{-1}$ for the epochs 2013, 2015, and 2017, respectively. Here, we considered $\sim$ 27 keV plasma and Galactic NH of 1.27$\times$$10^{21}$ cm$^{-2}$. For the epochs 2010 and 2019, the unabsorbed X-ray fluxes were found to be variable. During the epoch 2019, the flux was found to decrease by $\sim$ 22\% from the epoch 2010. The best-fit spectral parameters derived from the {\it Swift}/XRT observations are summarized in Table \ref{tab:swiftspecparam}.


\subsection{Phase-resolved spectroscopy} \label{sec:spinphasespec}
Using contemporaneous {\it Swift}/XRT and {\it NuSTAR} (FPMA and FPMB) observations, phase-resolved spectroscopy was also performed to trace the dependence of X-ray spectral parameters during the rotation of the WD. For this, we simply divided the spin pulse into ten equally spaced phase intervals from 0 to 1 in the phase interval of 0.1 by applying phase filters on the barycentric-corrected event file in {\sc xselect} package. The spectra of each phase segment were fitted with the best-fit model $D$, keeping fixed the plasma temperature and abundance values of the {\sc mkcflow}, and the parameters of the blackbody components at the values obtained from the average spectral fitting. Resulting spectral parameters are summarized in Table \ref{tab:xrayspinprs}. The free parameters are pwab components ($N_{H,max}$ and $\beta$), the normalization of the {\sc mkcflow} model ($n_{mkcflow}$), and the normalization of the Gaussian component ($n_g$). The significance of the variation of all of the components was measured using the $\chi^2$-test. We found that the variations of $N_{H,max}$, $n_{mkcflow}$, and $f_{X}$ are significant above a 90\%\ confidence level. Variations of these parameters are shown in Figure \ref{fig:xrayspinprs}. However, for the parameters $\beta$ and $n_{g}$, we did not find significant variations along with the spin phases. The variation of the absorption component ($N_{H,max}$) was found to be anti-correlated with the X-ray flux. However, an opposite pattern is seen in the variations of the X-ray flux and {\bf $n_{mkcflow}$}, that is, the variation of {\bf $n_{mkcflow}$} is correlated with the variation of the X-ray flux.


\section{Discussion} \label{sec:diss}
We have carried out a detailed X-ray analysis of an intermediate polar IGR1654. From the {\it NuSTAR} and {\it Swift} observations, we derived two significant periods, $P_\omega$ and $P_3$$_\omega$ in the 3-78 keV and 0.3-10.0 keV energy bands for the epochs 16 March 2019 and 08 March 2019, respectively. The periods derived for these epochs are well consistent (see Table \ref{tab:xrayps}). A significant spin period was also derived for the epoch 15 and 17 March 2019 of the {\it Swift} observations in the 0.3-10.0 keV energy band, which is similar to the period derived for the epochs described above. The derived X-ray spin period matches well with the optical spin period reported by \cite{Scaringi11}. The strong X-ray spin signal at $\sim$ 546 s is seen at all X-ray energies is characteristic of IPs and confirms the nature of the object. With the presence of optical beat ($\omega$$-$$\Omega$) and spin frequencies, \cite{Scaringi11} reported that IGR1654 displayed a mixture of disk and stream accretion. In contrast to \cite{Scaringi11}, we did not detect the beat or orbital frequencies in the X-ray power spectrum of IGR1654. The dominance of the spin frequency and the lack of beat frequency in the X-rays indicates that the system is accreting predominantly via a disk \citep[for details, see,][]{Mason88, Hellier91, Wynn92, Norton93, Norton96}. Although the existence of the beat modulation is not seen in the X-ray power spectrum of IGR1654, the spin modulations during an orbital cycle (Figure \ref{fig:orbspinprs}) do have, however, substantial implications for its accretion geometry. The observed spin modulations in IGR1654 might be affected due to absorption or occultation by a structure fixed in the binary frame or due to reprocessing of the spin pulse from the secondary star, the disk-stream region, or the inner Lagrangian point \citep{Norton92}.

Energy dependence of the spin pulse profile is one of the unique properties of IPs. The observed energy-dependent rotational modulations and hardening at the minimum intensity of the pulse profile as obtained from the {\it NuSTAR} and {\it Swift} data in the low energy ($<$ 10 keV) range can be attributed to the accretion curtain scenario and generally explained with variable complex absorbers \citep{Rosen88}. In this scenario, the accreting material falls onto the WD poles from the inner edge of the truncated disk at the magnetospheric radius and flows along the magnetic field lines in an arc-shaped curtain. In this model, the optical depth of the infalling material in the curtain is larger along the field lines rather than perpendicularly. Thus, photoelectric absorption is maximum when the accretion curtain points toward the observer and provides minimum spin modulations, and vice versa. No significant variations seen in the hardness ratio along with the spin phases above 10 keV imply that the accretion curtain scenario is not expected to produce significant spin modulations above 10 keV. \cite{Mukai99} and \citet{Demartino01} suggested that either reflection or tall shocks (shock height $>$ 0.1 $R_{WD}$) are responsible for the hard X-ray spin modulations in IPs. If reflection is the main mechanism for the hard X-ray modulations, then it should provide an anti-phased spin pulse profile with respect to a low energy pulsation. This is contrary to what we observe for IGR1654 (see Figure \ref{fig:nuflc}). On the other hand, the self occultation of tall shocks would provide the same phase spin modulations at both high and low energies and both accreting poles would be visible over a range of viewing geometry. To have the same-phased spin modulations at both hard and soft energies, $i$ + $\delta$ $<$ 90$^{\circ}$, where $i$ is the binary inclination and $\delta$ is the magnetic colatitude \citep{Mukai99}. This would suggest that the upper pole dominates the hard X-rays at all phases. In a few cases, where the hard X-ray amplitudes are observed on the order of 10 \% or less, the complex absorbers are found to be responsible for their modulations via Compton scattering \citep{Rosen92}. For IGR1654, the observed strong (29 \%) hard X-ray spin modulations are in the same phase with the low energy spin modulations, indicating that the hard X-ray spin modulations can be attributed to tall shocks above the accreting poles. Moreover, the height of the shock can be estimated using the reflection amplitude as described by \cite{Tsujimoto18}. If the reflector subtends $2\pi$ steradian of the sky, as seen from the X-ray emitter, then it provides a negligible shock height and only one pole is visible at all phases. However, when both emission regions are simultaneously observable and the value of the reflection amplitude is relatively small, then it implies a large shock height. Present X-ray data did not allow us to constrain the reflection amplitude (see Section \ref{sec:simspec}). Because of this, we were unable to constrain the shock height, but the high energy ($>$ 10 keV) spin modulation suggests that it is not small. Such hard X-ray modulations were also observed in the other two short-rotating period IPs V709 Cas and V2731 Oph, where a finite shock height was proposed as a solution for the spin modulations above 10 keV \citep{Demartino01, Mukai15, Lopes19}.

Following \cite{Norton99}, the observed double-humped pulse profile for a short rotating period ($P_\omega$$\sim$ 546 s) of the WD in IGR1654 can be explained with the two-pole accretion model. Two-pole accretion is believed to be a "normal" mode behavior in disk-accreting IPs, which can produce both either single-peaked or double-peaked pulse profiles, depending on the strength of the magnetic field of the WD and the visibility of two accreting poles during rotation of the WD \citep[see][]{Hellier96, Allan96}. \cite{Norton99} suggested two processes to produce a double-peaked spin pulse profile in the two-pole accretion scenario, which are briefly described below.

First possibility invokes large accretion areas due to the weak magnetic field of the short period rotating WD. The radius at which material is captured by the field lines is relatively small due to the low magnetic field of the WD. Consequently, the vertical optical depth is lower than the horizontal optical depth. Thus, when the upper pole is moving toward the observer, the minimum attenuation of the X-ray flux occurs and maximum emission is seen from it and gives the first peak in the pulse profile. At this phase, the lower pole is generally occulted since this pole is in anti-phase with that from the upper pole. However, when the upper pole is moving away from the observer, the lower pole is at its most visible, giving a second peak in the pulse profile.

Second possibility invokes tall accretion regions, but with the vertical optical depths being larger than the horizontal optical depths, as in the classical accretion scenario. The shock height is proportional to the size of the accreting area \citep{Frank92}, so the accretion regions in the short period rotating IPs are tall due to a weak magnetic field of the WD. When the upper pole points away from the observer giving maximum flux, the lower pole is viewed essentially from the side and its flux adds to the first maximum. However, when the upper pole points toward the observer giving minimum flux, the lower pole may still be visible and give rise to a second maximum. \cite{Norton99} also reported that slow rotators do not exhibit X-ray beat periods and they are an indicator of a WD with a relatively weak magnetic field. For IGR1654, we did not detect the X-ray beat signal in its power spectra; also, the spin modulations are double-humped where high energy spin modulations suggest that it has a tall shock. Therefore, the second possibility seems to be the most feasible to produce a double-humped spin pulse profile in a short period IP IGR1654. In contrast to {\it NuSTAR}, triple-hump profiles of the {\it Swift} data for the epoch 08 March 2019 are difficult to explain using the two-pole accretion model. To accommodate for them in a two-pole accretion model, two adjacent peaks with the presence of a narrow dip (at a spin phase $\sim$ 0.3) could represent the emission from one dominant pole, whereas a relatively smaller amplitude hump after a wider dip (at spin phase $\sim$ 0.6-0.7) may represent the emission from the second pole. If two nearby peaks represent one dominant pole, then the pole should remain in the line of sight for a longer time. This is also evident from the observed pulse profile of IGR1654 in which the dominant pole remains in view for almost half of the spin cycle. A narrow dip between two adjacent peaks is not prominent in all energy bands, indicating that the possibility of absorption is more likely the cause for the appearance of the triple-hump pulse profile. Some short-period IPs, such as V405 Aur, AE Aqr, YY Dra, DQ Her, XY Ari, GK Per, V709 Cas, V667 Pup, V515 And, and V2731 Oph \citep[see][]{Allan96, Hellier96, Kamata93, Norton88, Norton99, Butters07, Butters08, Martino08, Lopes19}, have shown double-peaked X-ray spin pulse profiles on some occasions and they are explained with the two-pole accretion scenario. Also, these short-period IPs did not exhibit clear X-ray beat periods in their lives. The absence of a beat signal and observed double-humped pulse profile at a spin period of $\sim$ 546 s from two-pole accretion suggests that IGR1654 can also be placed in the list of the short-period IPs described above. In contrast to X-rays, \cite{Scaringi11} detected a single-peaked and quasi-sinusoidal spin modulation in the optical domain for IGR1654. This seems to be very interesting since the X-ray spin pulse profile is double-peaked for this system. This can be interpreted with the standard accretion-curtain model. In this model, optical emission originates from the accretion curtains between the inner disk and the WD. If they are optically thick, their varying aspect produces modulations during WD rotation. In this case, both poles of the WD act in phase due to which the changing visibility of the curtains results in a single-peaked pulse with a maximum when the upper pole points away from the observer \citep{Hellier91, Hellier95, Kim96}.

The average X-ray spectrum of IGR1654 is well modeled by a complex absorber with a maximum equivalent hydrogen column of $\sim$ 1.8$\times$$10^{23}$ cm$^{-2}$ and a power-law index of -0.22 for a covering fraction, a multi-temperature cooling flow model at temperatures 0.0808 keV (fixed) and $\sim$ 31 keV, along with a blackbody with an average temperature of $\sim$ 64 eV. In contrast to other soft-IPs \citep[see][]{Mason92, Haberl94, Haberl02, Evans04, Martino04, Vrielmann05, Anzolin08, Joshi16, Joshi19}, the evidence of a soft X-ray excess is meager in the X-ray spectrum of IGR1654. If the soft X-ray excess is really present in IGR1654, then it can be due to the heated region near the accretion footprints which is not hidden by the accretion column depending upon the system inclination and the magnetic colatitude \citep[see][]{Evans07}. We also derived unabsorbed soft ($F_s$) and hard ($F_h$) X-ray fluxes of 5.02$_{-6.90}^{+3.55}$$\times$$10^{-12}$ erg cm$^{-2}$ s$^{-1}$ and 3.00$_{-0.03}^{+0.03}$$\times$$10^{-11}$ erg cm$^{-2}$ s$^{-1}$ in the energy band of 0.5-78.0 keV from the blackbody and mkcflow models, respectively. The softness ratio, $F_s$ /4$F_h$, was then calculated as $\sim$ 0.04 for IGR1654, which seems to be matched with the softness ratios observed for other soft-IPs \citep{Evans07}. We also determined the size of the accretion footprint from the soft X-ray flux. The unabsorbed soft X-ray flux of 5.02$\times$$10^{-12}$ erg cm$^{-2}$ s$^{-1}$ and the temperature of 64 eV imply an emitting area of 3.95$\times$$10^{13}$ cm$^{-2}$. The WD mass of 0.74$_{-0.08}^{+0.09}$ $M_\odot$ in IGR1654 implies the observed blackbody emitting area covers $\sim$ 6$\times$10$^{-4}$ of the WD surface, which is consistent with other estimates for the accretion area in IPs \citep{Hellier97}. The above discussion is purely based on the presence of a soft X-ray excess only in the X-ray spectra of IGR1654.

From spin-phase-resolved spectroscopy, we have found that the variation in the absorption component is anti-correlated with the X-ray flux, that is, the absorption component is found to be maximum when the X-ray flux is minimum. Such variations can be explained with the widely accepted classical curtain scenario. During the rotation of the WD, when the curtain points toward the observer, the X-ray flux is minimum due to the maximum absorption, however, the X-ray flux is maximum when the accretion curtain moves away from the line of sight of the observer.


\section{Summary} \label{sec:sum}
To summarize, the detection of the X-ray spin period at $\sim$ 546 s is well consistent with the previously obtained optical spin period and unambiguously confirms the IP nature of this system. The presence of the strong spin pulse, with no sign of orbital or side-band periodicities in the X-rays, indicates that the system is accreting predominantly via a disk. A variable covering absorber appears responsible for the spin pulsations in the low energy range, however, the modulations above 10 keV can be attributed to tall shocks. The observed double-humped X-ray pulse profiles at low and high energies reveal two-pole accretion geometry with tall accretion regions in short rotating IP IGR1654, which is probably caused by a weak magnetic field of the WD. The X-ray spectrum of IGR1654 is affected by complex absorption with an equivalent hydrogen column of $\sim$ 1.8$\times$$10^{23}$ cm$^{-2}$ and a power-law index of -0.22 for a covering fraction, as well as described by a soft X-ray blackbody component at $\sim$ 64 eV and an optically thin plasma emission component at $\sim$ 31 keV. Variations in the absorption component and X-ray flux along with the WD rotation are compatible with the classical accretion curtain scenario.

\section*{Acknowledgements}
We acknowledge the referee for useful comments and suggestions that improved the manuscript considerably. This research has made use of data obtained with the {\it NuSTAR} mission, a project led by the California Institute of Technology (Caltech), managed by the Jet Propulsion Laboratory (JPL) and funded by NASA. This research has made use of the XRT Data Analysis Software (XRTDAS) developed under the responsibility of the ASI Science Data Center (ASDC), Italy. This work is supported by the National Program on Key Research and Development Project (Grants No. 2021YFA0718503, 2021YFA0718500), and the NSFC (12133007, U1838103, 11622326). One of the authors Ashish Raj acknowledges the Research Associate Fellowship with order no. 03(1428)/18/EMR-II under Council of Scientific and Industrial Research (CSIR).

\bibliographystyle{aa}
\bibliography{ref}
 
\end{document}